# Theoretical study of small signal modulation behavior of Fabry-Perot Germanium-on-Silicon lasers

Ying Zhu[1,2], Liming Wang[1], Zhiqiang Li[3], Ruitao Wen[2], and Guangrui Xia[1]

*1 Department of Materials Engineering, the University of British Columbia, Vancouver BC, V6T 1Z4 Canada*

*2 Department of Material Science and Engineering, Southern University of Science and Technology, Shenzhen 518055, China*

*3 Crosslight Software Inc., Vancouver, BC V5M 2A4, Canada*

**Abstract:**
This work investigated the small signal performance of Fabry-Perot Ge-on-Si lasers by modeling and simulations. The 3dB bandwidth dependence on the structure parameters such as poly-Si cladding thickness, Ge cavity width and thickness, and minority carrier lifetime were studied. A 3dB bandwidth of 33.94 GHz at a biasing current of 270.5 mA is predicted after Ge laser structure optimization with a defect limited carrier lifetime of 1 ns.

## 1. Introduction

Optical communication utilizes light to carry and transport information, which possesses a large transmission bandwidth, fast communication speed and a low transmission loss compared to metal interconnects. In short-reach communication, optical communication has also been playing an increasingly significant role in applications such as backbone interconnect in data centers. For communication of mm and shorter distance, such as on-chip communication, silicon photonics have been considered as a powerful tool to address the slow-speed and high energy consumption of metal interconnect [1-5].

Although most silicon-integrated photonic components are mature, Si-compatible lasers have been sought for decades [6-10]. III-V semiconductor materials have been widely used as active laser gain mediums, and III-V semiconductor lasers integrated on Si substrate by heteroepitaxy [11, 12] or bonding [6, 13, 14] techniques have showed very good performance up to now. Even though the direct heteroepitaxial growth of III-V quantum dot (QD) lasers on Si substrates has been demonstrated to have great potentials [15-17], the unavoidable contamination problems impede the entrance of III-V semiconductors into the mainstream Si fabrication facility. In recent two decades, germanium (Ge), a group IV element semiconductor, has been demonstrated as a promising gain medium material. Moreover, compared with III-V semiconductors, the fabrication of Ge is more compatible with the mainstream Si fabrications processes, which has great benefits in realizing Si-compatible lasers [18-21].

Ge is an indirect bandgap (0.664 eV) material, but only has a small energy difference (136 meV) between the direct valley (Γ) and indirect valley (L) [22]. Meanwhile, the direct bandgap of Ge is 0.8 eV, corresponding to a wavelength of 1550 nm, which exactly lies in the low loss window of Si dioxide. In 2007, Jifeng Liu et al. have theoretically predicted that Ge can achieve an optical gain >

500 cm$^{-1}$ and a relatively high differential gain of $8.0 \times 10^{-16}$ cm$^2$ under an injected carrier density of 2.0 to $4.0 \times 10^{17}$ cm$^{-3}$ by combining the strain-induced bandgap engineering and n-type doping above $4.0 \times 10^{19}$ cm$^{-3}$ [7, 23]. Based on that, an optically pumped edge-emitting Ge-on-Si laser has been successfully demonstrated with a gain spectrum in 1590-1610 nm in 2010 [24]. Later in 2012 [25] and 2015 [26], electrically pumped Fabry-Perot Ge-on-Si lasers were realized. However, these early Ge-on-Si lasers had a high threshold current density of 280 kA·cm$^{-2}$ and low wall-plug efficiency of 0.5-4% [25, 27]. In comparison, commercial III-V lasers have low threshold current densities of 0.01~1 kA·cm$^{-2}$ and high wall-plug efficiencies in the range from 50% to 65%. Some theoretical studies were available to calculate the performance potential of Ge lasers, including 0-dimentional (0D) analysis of Ge optical gains and their strain and doping dependences [28]. 2D Ge-on-Si laser modeling and simulation studies from our group [29] predicted that Ge-on-Si lasers could be greatly improved to reach a wall-plug efficiency of 43.8% and a threshold current of 4 mA (current density of 3 kA·cm$^{-2}$) by structure optimization, strain engineering and Ge material improvement. In 2016 and 2017, lasers based on Ge nanowires under 1.6% uniaxial tensile strain were designed and fabricated, which achieved a pulsed lasing with a low optical pumping threshold density of 3.0 kW·cm$^{-2}$ [30, 31].

In optical links, direct modulated lasers (DMLs) are desired, which eliminates the needs for external modulators in photonic circuits. For DMLs, dynamic properties, such as small signal behaviors are crucial, which provide the information of the maximum frequency, modulation bandwidth that lasers can be modulated. So far, Ge lasers research has been focused on the static properties, such as optical gains and losses, threshold current densities and efficiencies. It is important to investigate the direct modulation potentials of Ge lasers, and the methods to improve the bandwidth of Ge lasers with insights gained from III-V semiconductor lasers [32-35].

This work is a 2D theoretic study of Ge-on-Si small signal modulation behaviors. We have investigated the small signal modulation responses of Ge-on-Si lasers, and how modulation bandwidth can be improved by structure optimization and material quality enhancement.

**2. Ge laser modeling and calibration**

2D Ge-on-Si laser modeling was conducted and calibrated in our group's previous studies in [29, 36], where models of Ge energy band structure (with biaxial strain and doping) and material loss (including the bandgap narrowing effect and the energy separation effect) were implemented in LASTIP$^{TM}$, a mainstream commercial 2D laser simulation software. Due to the software updates, some small changes in the best fitting parameters were expected in the 2020 version that was used in this work. Due to the lack of experimental data, the laser structure and L-I curve in this work are also based on the MIT's electrically pumped Ge-on-Si laser experiment [25]. The Fabry-Perot laser structure model is shown as Figure 1. The thickness, doping, and strain parameters of each layer material are the same as those in the experiments in Ref. [25], as shown in Table I.

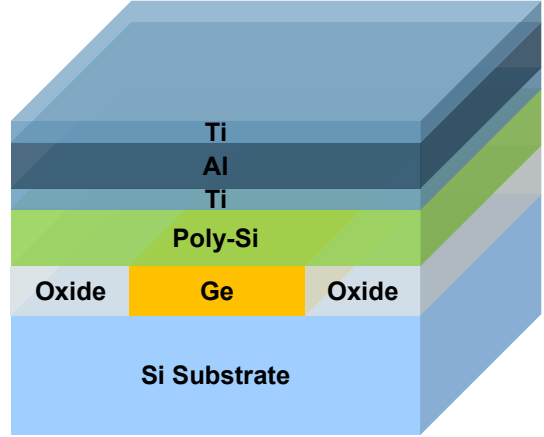

Figure 1. The simulated Ge-on-Si laser structure based on the early work at MIT in Ref. [25].

Table I The doping and strain parameters of Ge-on-Si lasers.

| | Doping type | Doping concentration (cm$^{-3}$) | Strain | Thickness |
|---|---|---|---|---|
| Si substrate | n-type | $5 \times 10^{19}$ | - | 2 μm (to reduce the computing time) |
| Ge | n-type | $4 \times 10^{19}$ | 0.25% biaxial tensile strain | 200 nm |
| Poly-Si | p-type | $3.6 \times 10^{20}$ | - | 180 nm |

To reduce the computation time, 2 μm thick Si substrate was used, and a virtual contact was set at the bottom of Si substrate and at the top of the metal layers for the biasing purpose. The active Ge region has a width of 1 μm, a length of 270 μm long and a thickness of 200 nm which was the average value of the 100 to 300 nm thickness from the experiment results owing to the process non-uniformity [25].

The metal-semiconductor heterojunctions were aligned by electron affinity. The reflectivity values of the two facets are $R_1 = 23\%$ and $R_2 = 38\%$, which corresponds to a mirror loss $\alpha_m$ of 45 cm$^{-1}$ based on the MIT work. The refractive index values of all materials are wavelength dependent. These material parameters mainly come from literature [29, 36] and [27]. The Auger coefficients are set as $C_{nnp} = 3.0 \times 10^{-32}$ cm$^6 \cdot$s$^{-1}$ and $C_{ppn} = 7.0 \times 10^{-32}$ cm$^6 \cdot$s$^{-1}$ [23]. Due to the lack of the relevant experiment data of Ge thin films, surface recombination was not considered in our simulations here. The defect limited lifetime ($\tau_{n,p}$) is an important material quality parameter and has been estimated conservatively to 1 ns for the epitaxial grown Ge films, based on the measurements in recent work [37].

As for the laser's performance, the loss will be one of the most essential influenced factors. Generally, the optical loss is composed of two loss mechanisms: the internal loss $\alpha_i$ and the mirror loss $\alpha_m$. Here the internal loss $\alpha_i$ is assumed to be dominated by the free carrier absorption (FCA) [7]. In LASTIP$^{TM}$, for a narrow wavelength range, the free carrier absorption (FCA) is described by

$$\alpha_i = AN + BP, \quad (1)$$

where A and B are constants, N and P are the electron and hole density, respectively, in the unit of cm$^{-3}$. We have used the first principle calculations of free carrier absorption results in the n-type Ge [27] and experiment measurements in the p-typed Ge [38] as a starting point and obtained the best fitting result to the L-I curve with the following free carrier absorption relationship:

$$\alpha_i = 4.18 \times 10^{-19} N + 1.021 \times 10^{-17} P, \quad (2)$$

The fitting values were slightly different from our previous reported best-fitting parameters $\alpha_i = 5 \times 10^{-19} N + 1.023 \times 10^{-17} P$ [29] because of the updates in the new version of the simulation software. The free carrier absorption relationship in the doped Si substrate and poly-Si cladding layer were obtained from work [39] and [40].

As the data from MIT's Fabry-Perot lasers were measured from one side of the laser, we used the laser optical power from the polished facet ($R_1$ = 23%) as the optical output power to fit. Based on all these material parameters, our simulation models provided a current density $J_{th}$ of 297.3 kA·cm$^{-2}$ or threshold current $I_{th}$ of 802.7 mA at 15°C with the transverse electric (TE) mode lasing at λ = 1676 nm. This fitted result was very closed to the MIT experiment results of 280 kA·cm$^{-2}$ and lasing wavelength of 1650 nm [25], which can be seen in the Figure 2.

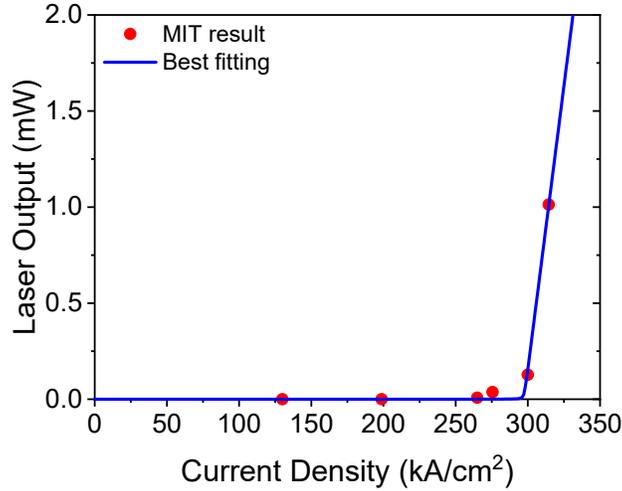

Figure 2. L-I curves from the MIT experiment in Ref. [25] result and our best-fitting curve.

3. **Relevant theories to calculate the frequency responses**

After fine-tune the best-fitting parameters in the 2020 version LASTIP$^{TM}$, we were ready to calculate the small signal modulation properties. First, let's review the relevant theories on that.

3.1 Frequency calculations

The relaxation resonance frequency is given by [41]

$$f_r = \frac{1}{2\pi} \sqrt{\frac{v_g}{q} \Gamma \eta_i \frac{dg}{dN} (I_b - I_{th})/V}, \quad (3)$$

where $v_g$ is the group velocity, $q$ is the elementary charge, $\Gamma$ is the optical confinement factor, $\eta_i$ is the internal efficiency, $dg/dN$ is the differential gain, $I_b$ is the biased current, $I_{th}$ is the threshold current, and $V$ is the volume of active region. When the damping factor is small or negligible, the electrical 3dB down frequency is given by [41]

$$f_{3dB} = \sqrt{1+\sqrt{2}} f_r \approx 1.55 f_r, \quad (4)$$

Based on the equation (1), there are mainly four strategies to enhance the modulation frequency: to improve the confinement factor; to minimize the active region volume; to enhance the differential gain, to maximize the difference between the biased current and the threshold value; and to enhance the internal efficiency.

3.2 Differential gain of strained Ge with doping

Optical gain, $g$, is another important parameter for the active medium materials, which determines the capability of laser medium to increase the output power. The differential gain, $dg/dN$, is a critical parameter in high-speed laser applications, because the relaxation resonance frequency of the laser depends on the square root of the differential gain [42], shown as equation (3).

The optical gain, $g$, is related to the carrier density, $N$, by a simple two parameter logarithmic formula [41], shown in equation (5):

$$g = g_{0N} \ln \frac{N}{N_{tr}} \quad (g > 0), \quad (5)$$

where $g_{0N}$ is the gain coefficient, $N_{tr}$ is the transparency carrier density. Under this condition, the differential gain ($dg/dN$), which is of great importance in our simulation and directly influenced by the carrier density and gain coefficient, is given by:

$$\frac{dg}{dN} = \frac{g_{0N}}{N}, \quad (6)$$

Above threshold, the carrier density, $N$, equals to the threshold carrier density, $N_{th}$. The optical gain behavior of 0.25% tensile strained n-doped Ge has been modeled and simulated by Jifeng Liu et. al, [23]. They have demonstrated that a significant net gain of about 400 cm$^{-1}$ can be achieved in the 0.25% tensile strained n-doped Ge with an extrinsic electron density of $7.6 \times 10^{19}$ cm$^{-3}$. A high differential gain of $dg/dN = 8 \times 10^{-16}$ cm$^2$ could be obtained at a relatively low injected carrier density. Here using our calibrated Ge models in LASTIP$^{TM}$, we have showed the net modal gain in 0.25% tensile strain n-doped of $4 \times 10^{19}$ cm$^{-3}$ Ge versus wavelength at various biasing current ranging from 0 to 999 mA in Figure 3(a). From that we can see the net modal gain became positive from the biasing current of 432 mA. Finally, it reaches 45 cm$^{-1}$ net modal gain at the threshold current of 802.7 mA, which equals to the mirror loss, the onset of lasing. The peak material gain and differential gain at the peak of the gain curve, $dg/dN$, versus carrier density for the 0.25% tensile strained Ge with a doping concentration of $4 \times 10^{19}$ cm$^{-3}$ were calculated and plotted in Figure 3(b). We can see that the transparency carrier density is $3.1 \times 10^{19}$ cm$^{-3}$ and the peak material gain increases with the carrier concentration [23] and could reach as high as 2034.9 cm$^{-1}$ at the carrier density of $9.44 \times 10^{19}$ cm$^{-3}$. From the differential peak gain versus the carrier density, there is a great peak appearing near the point of material gain becoming positive, which showed a relatively high differential gain value of $1.35 \times 10^{-16}$ cm$^2$ at a carrier density of $3.5 \times 10^{19}$ cm$^{-3}$. After that, the differential gain decreases dramatically and then reaches a plateau with the increase of carrier concentration [43, 44]. From that and based on the equation (6), we can see that the differential gain is greatly dependent on the carrier concentration. Hence when designing high-speed lasers, it is significant to let lasers working closer to the transparency state to obtain high differential gain, which means letting threshold carrier density getting close to transparency carrier density.

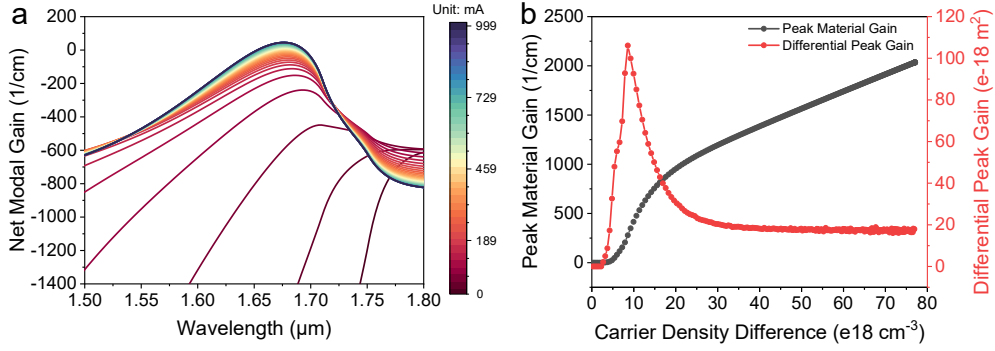

Figure 3. The optical gain properties of 0.25% tensile strain n-doped Ge (N = 4 × 10$^{19}$ cm$^{-3}$). a) The net modal gain versus wavelength under different biasing current (from 0 to 999 mA). b) Peak material gain and differential peak gain $dg/dN$ (at the peak of the gain curve) versus carrier density.

3.3 Laser output calculations

Based on the laser rate equations, the optical laser output power, $P_{out}$ was calculated as a function of the biasing current, $I$ [41].

$$P_{out} = \eta_i \frac{\alpha_m}{\langle\alpha_i\rangle+\alpha_m} \frac{h\nu}{q}(I - I_{th}) = \eta_d \frac{h\nu}{q}(I - I_{th}), \tag{7}$$

$$\eta_d = \frac{\Delta P}{\Delta I} / \frac{hc}{q\lambda} = \eta_i \frac{\alpha_m}{\langle\alpha_i\rangle+\alpha_m} = \eta_i \eta_{ext}, \tag{8}$$

The differential efficiency $\eta_d$ is defined as the product of internal efficiency $\eta_i$ and the extraction efficiency $\eta_{ext}$, shown as equation (8). The extraction efficiency $\eta_{ext}$ is defined by equation (9), in which the $\langle\alpha_i\rangle$ is the weighted average of the local loss.

$$\eta_{ext} = \frac{\alpha_m}{\langle\alpha_i\rangle+\alpha_m}, \tag{9}$$

where $\frac{\Delta P}{\Delta I}$ is the slope of the light-current (L-I) curve, $h$ is the Plank constant, $c$ is the speed of light, $q$ is the elementary charge, $\lambda$ is the lasing wavelength, $I$ is the biased current, $I_{th}$ is the threshold current.

The threshold current, $I_{th}$, and threshold carrier density, $N_{th}$, are expressed in the following equations [41]:

$$I_{th} = \frac{qdWL}{\eta_i}(R_{SRH}(N_{th}, P_{th}) + R_{Rad}(N_{th}, P_{th}) + R_{Auger}(N_{th}, P_{th})) = \frac{qdWL}{\eta_i}\frac{N}{\tau_c}, \tag{10}$$

$$N_{th} = N_{tr} + \frac{\langle\alpha_i\rangle+\alpha_m}{\Gamma(d,W)G'}, \tag{11}$$

where $q$ is the elementary charge, $d$, $W$, $L$ are the thickness, width, and length of the active region, respectively, $\eta_i$ is the internal efficiency, $R_{SRH}$ is Shockley-Read-Hall non-radiative recombination rate generating at defects [45], $R_{Rad}$ is the spontaneous recombination rate, $R_{Auger}$ is the nonradiative recombination rate due to the Auger recombination process, $\tau_c$ is the carrier lifetime, $N_{tr}$ is the transparency carrier density, $\Gamma$ is the optical confinement factor which is influenced by the thickness $d$ and width $W$ of the active region, $G'$ is the material gain, which equals to $G/\Gamma$.

3.4 Model assumptions and limitations

In our modeling, we had some assumptions and conditions to make the calculation more reasonable and simpler. First, with the high pumping current, high heat generation is expected that can cause electron migration and contact metal melting. In this work, these thermal effects have not been considered here and the temperature of the lasers simulated was set at 288 K. This condition can be satisfied when cooling is used in the laser measurements [25]. Second, LASTIP$^{TM}$ is a 2D simulator, which ignores the phase matching condition and assumes that only a single longitudinal mode exists, and that lasing occurs at a wavelength with the peak modal gain [46]. In a real Fabry-Perot laser, lasing is at the wavelength where the cavity round-trip gain peaks. Third, the gain saturation effect was included here, as we assumed that in our simulation such high photon density would not encounter. Theoretically, gain saturation will reduce the differential gain and then affect the resonance frequency [47]. As the carrier concentrations are not high compared to quantum-well or QD lasers with relatively small active volume, the gain saturation effect is less important in the Fabry-Perot lasers studied here [41].

3.5 Small signal response simulated without laser structure optimization

First, the small-signal-modulation properties of the laser structure from the MIT work was calculated by LASTIP$^{TM}$ (Figure 1), under a biased current ranging from 756 to 999 mA, shown in Figure 4. This current range was selected to cover a current window from below to above the threshold current of 802.7 mA. Under a simplified condition without considering the electron migration and metal heating effect of the metal contact, the 3dB bandwidth was calculated to be about 6.2 GHz at the maxim simulated biased current of 999 mA. This bandwidth value is relatively small, and the biased current is very high practical applications. Next, we investigated the impacting factors, such as the laser structure parameters and the minority carrier lifetime, to improve this modulation bandwidth.

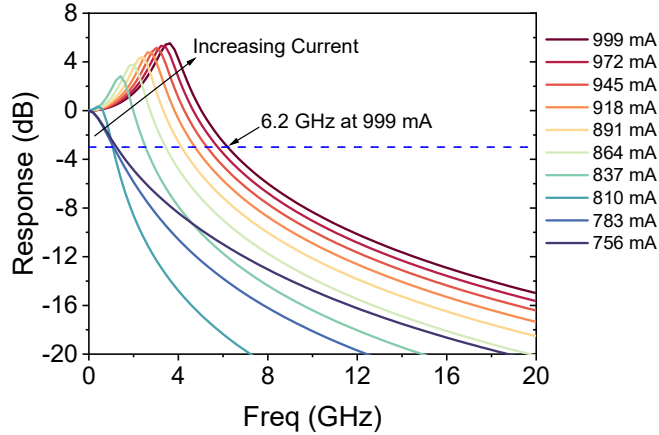

Figure 4. Simulated small-signal-modulation responses of the Fabry-Perot laser structure shown in Figure 1.

## 4. Optimization of Fabry-Perot Ge-on-Si lasers

To investigate these influencing factors and the small signal modulation responses of the Ge lasers,

we optimized the structure of the Ge laser. There are mainly three optimizing parameters, including the thickness of the poly-silicon cladding layer $d_{poly-Si}$, the width of Ge cavity $W_{Ge}$ and the thickness of the Ge cavity $d_{Ge}$. The length of Ge cavity was maintained at 270 μm. Based on the previous simulations [36], this length gives low threshold currents. Ge is not a well-understood optical gain material, and many model parameters do not have widely agreed values or ranges. Therefore, during our simulation process, our goal was not to obtain the ultimate optimal point and the exact values for these properties, but rather to demonstrate that the performance of Ge-on-Si lasers could be improved greatly and how each factor influences the small-signal-modulation responses.

After the calibration of our model parameters and analysis of the laser key performance equations, we started optimizing the Fabry-Perot Ge-on-Si lasers with LASTIP$^{TM}$. The starting point was the laser structure and parameters in Figure 1 and Table 1, where the active Ge region has a width of 1 μm, a length of 270 μm long and a thickness of 0.2 μm and the poly-Si thickness is 0.18 μm. For an ideal laser, small threshold current, high efficiency and large bandwidth are all desired properties, but they cannot be obtained at the same time. As we concentrate on studying the small signal modulation of lasers, the 3dB bandwidth was chosen as the most important optimization criteria. In the simulation, the biased current was set at $I_b$ = 270.5 mA through all the 3dB bandwidth calculation to keep only one variable at one time and make the geometry improvement simpler to view. The biased current value was chosen to be larger than the threshold currents of all geometry simulated, and not higher than 10 times of $I_{th}$.

### 4.1 Poly-Si thickness $d_{poly-Si}$ dependence

The thickness of the poly-Si layer has a significant effect on the optical internal loss, which directly impacts the differential efficiency and threshold current as shown in equation (8) and (11). From Figure 5(a), the calculated internal loss decreases significantly with thicker poly-Si cladding and then becomes steady. The extraction and differential efficiency increase and finally reach a plateau. This is because the metal contact has a much higher optical absorption loss than poly-Si. As the poly-Si becomes thicker, the top metal contact is moved further away from the Ge active region and the internal loss $\langle \alpha_i \rangle$ decreases [48]. Hence the internal efficiency increases a lot at first and then becomes steady. The optical confinement factor increases at first due to the metal part moved away, and the metal optical absorption reduces significantly. With the increase of the poly-Si thickness, the optical absorption in the poly-Si increases, which leads to a slight decrease of the optical confinement factor (Figure 5(b)).

Since the internal loss decreases, the threshold current exhibits a similar decreasing behavior and becomes steady eventually, as shown in Figure 5(c). Moreover, the threshold carrier density also decreases and then reaches a plateau since there is a smaller loss to compensate for. According to the differential gain in equation (6), the differential gain, $dg/dN$, displays an opposite trend with the threshold current, that is, increases fast first and then becomes steady (Figure 5(d)). Consequently, with the increasing of poly-Si thickness, the 3dB bandwidth increases to a maximum value of 27.08 GHz at the biased current of 270.5 mA at $d_{poly-Si}$ = 0.7 μm and then slightly decreases due to the minor reduction of confinement factor (Figure 5(d)). Therefore, 0.7 μm was chosen as the optimized poly-Si cladding layer thickness as it shows the highest bandwidth value.

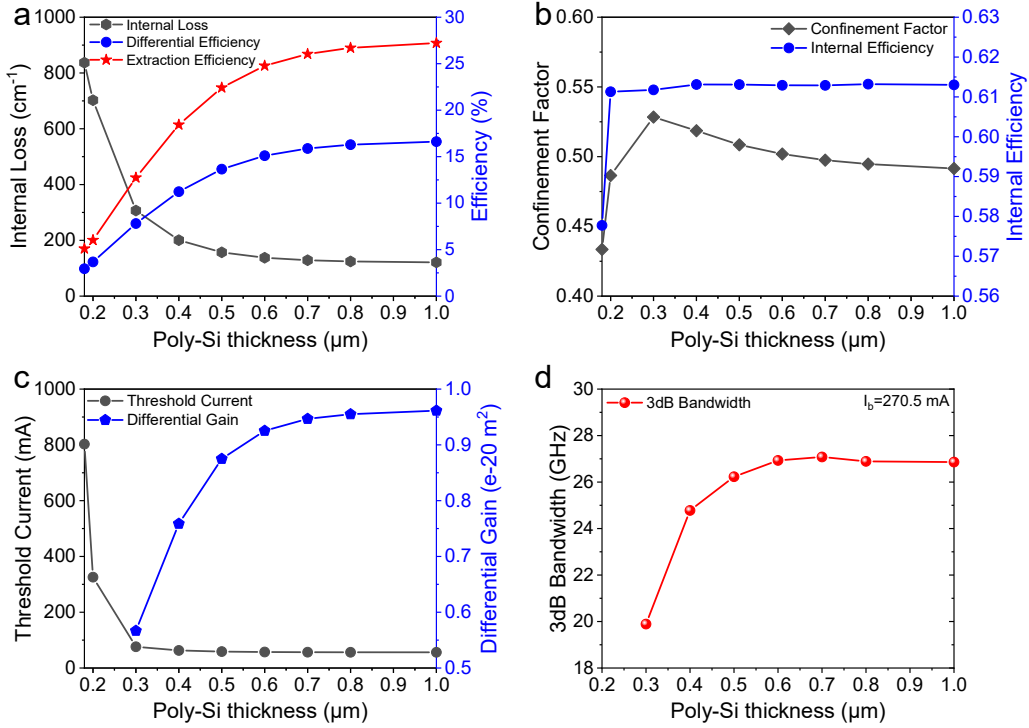

Figure 5. Poly-Si thickness dependence in the range of 0.18 to 1 μm ($W_{Ge}$ = 1 μm, $d_{Ge}$ = 0.2 μm). a) The internal loss, $\langle \alpha_i \rangle$, differential efficiency, $\eta_d$, and the extraction efficiency, $\eta_{ext}$. b) Confinement factor, $\Gamma$, and internal efficiency, $\eta_i$. c) Threshold current, $I_{th}$, and differential gain, $dg/dN$. d) The 3dB bandwidth at the biased current of 1002 A/m or 270.5 mA. This biased current value was chosen to be larger than the threshold current of all geometry changing range, and not higher than 10 times $I_{th}$ to guarantee the lasers working properly.

## 4.2 Ge width $W_{Ge}$ dependence

The width of Ge has a direct impact on the bandwidth through two parameters: (1) optical confinement factor, $\Gamma$, and (2) active region volume, $V$. The influence of the Ge width on $\langle \alpha_i \rangle$, $\eta_d$, $\eta_{ext}$, $\Gamma$, $\eta_i$, $I_{th}$, $V$, $dg/dN$ are shown in the Figure 6. With the increase of the Ge width, the internal loss becomes smaller, and the differential and extraction efficiency rise a little and then plateau as in Figure 6(a), because a wider Ge cavity results in a larger Ge active region and larger confinement factor in Figure 6(b), and the optical mode will less extend into the lateral layers. Meanwhile the internal efficiency shows very little decreasing tendency with wider Ge cavity because a narrower waveguide is beneficial for the uniform current injection. It is obvious that the volume of the active region is in a monotonically linear increasing relationship with Ge width, shown in Figure 6(c). Based on the equation (10) for threshold current, the threshold current is directly proportional to the width of Ge cavity and has an indirect relationship with internal loss by the threshold carrier density, $N_{th}$. Under the combined action of Ge width and internal loss, the threshold current exhibits a minimum value of 32.73 mA at a width of 0.5 μm and then increases greatly with wider Ge width. The variation tendency of theoretical threshold carrier density can be inferred from equation (11). With the decrease of internal loss and increase of confinement factor, the $N_{th}$ displays a decreasing tendency and then becomes steady. Hence, the variation trend of differential gain shows an increasing part at first and then plateaus. Combined all these competing factors and based on equation (3,4), the 3dB bandwidth climbs to a maximum value of 31.72 GHz at a width of 0.6 μm, and then reduces a lot due to the rapidly increasing active region volume. Although the 0.6 μm Ge width does not possess the lowest threshold current, it is a tradeoff to choose the optimization point between threshold current and 3dB bandwidth. Since the 3dB bandwidth is our final goal, the Ge width has been set at 0.6 μm to get the highest performance in modulation.

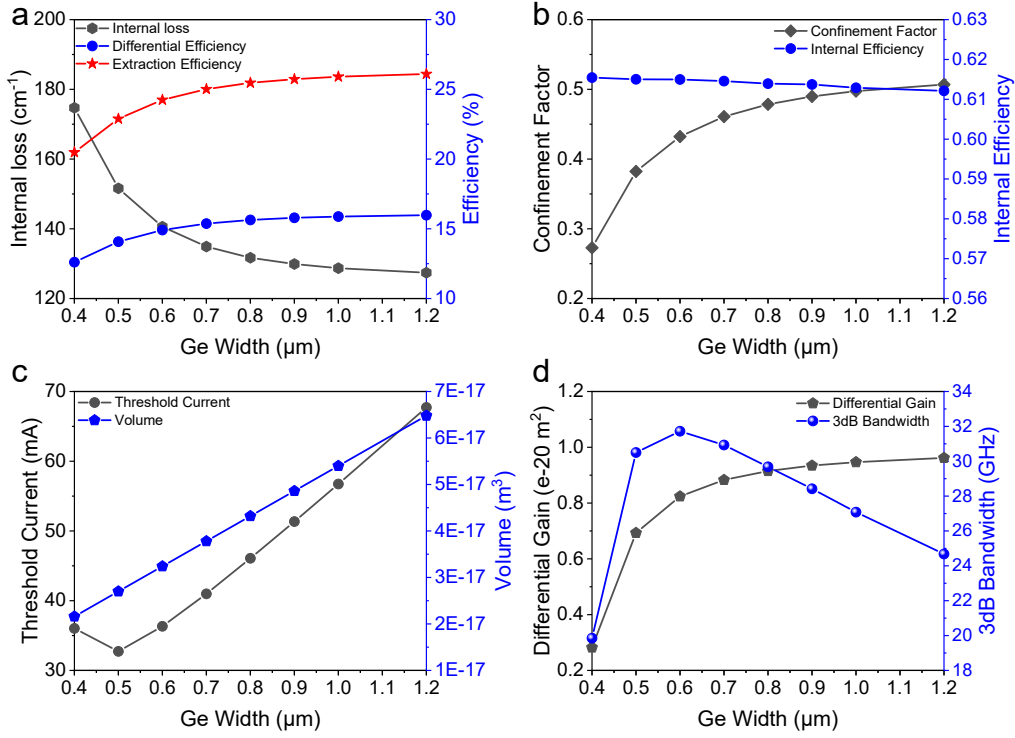

Figure 6. Ge width dependence in the range of 0.4 to 1.2 μm ($d_{poly-Si}$ = 0.7 μm, $d_{Ge}$ = 0.2 μm). a) The internal loss, $\langle\alpha_i\rangle$, differential efficiency, $\eta_d$, and the extraction efficiency, $\eta_{ext}$. b) Confinement factor, $\Gamma$, and internal efficiency, $\eta_i$. c) Threshold current, $I_{th}$, and volume of the Ge active region, $V$. d) The differential gain, $dg/dN$, and 3dB bandwidth at the biased current of 1002 A/m or 270.5 mA. This biased current value was chosen to be larger than the threshold current of all geometry changing range, and not higher than 10 times $I_{th}$ to guarantee the lasers working properly.

### 4.3 Ge thickness $d_{Ge}$ dependence

As for the thickness of Ge active region, it has a similar dependence as the Ge width dependence, (1) optical confinement factor, $\Gamma$, and (2) active region volume, $V$. The influence of the Ge width on $\langle\alpha_i\rangle$, $\eta_d$, $\eta_{ext}$, $\Gamma$, $\eta_i$, $I_{th}$, $V$, $dg/dN$ are shown in the Figure 7. The internal loss exhibits a decreasing trend because the thicker Ge layer leads to less mode overlapping with lossy poly-Si cladding and metal contact layers. Thus, the differential and extraction efficiency rise at the beginning and then become steady (Figure 7(a)). The internal efficiency shows minor decrease due to larger resistivity in Ge. The confinement factor increases dramatically with the thicker Ge layer which can provide better vertical confinement (Figure 7(b)). Also, the volume of active region is directly proportional to the thickness of Ge layer (Figure 7(c)). Besides, based on equation (10), the threshold current is influenced by Ge thickness, internal loss, and confinement factor. Under these competing actions, the threshold current is finally dominated by the thickness of Ge and monotonically increasing. With the increase of Ge thickness, the lasing wavelength has a red shift, and the refractive index becomes smaller, which influences the material gain coefficient and differential gain. The differential gain peaks at a Ge thickness of 0.3 μm and then decreases to a plateau (Figure 7(d)). Based all these influenced aspects, the 3dB bandwidth shows a peak of 33.94 GHz at a thickness of 0.3 μm and then decreases greatly, which is dominated by the increase of the active region volume.

Above all, we have come to an optimization point of the Ge-on-Si lasers, which is $d_{poly-Si}$ = 0.7 μm, $W_{Ge}$ = 0.6 μm and $d_{Ge}$ = 0.3 μm, with a 3dB bandwidth of 33.94 GHz at a biased current of 270.5 mA and a threshold current of 46.42 mA.

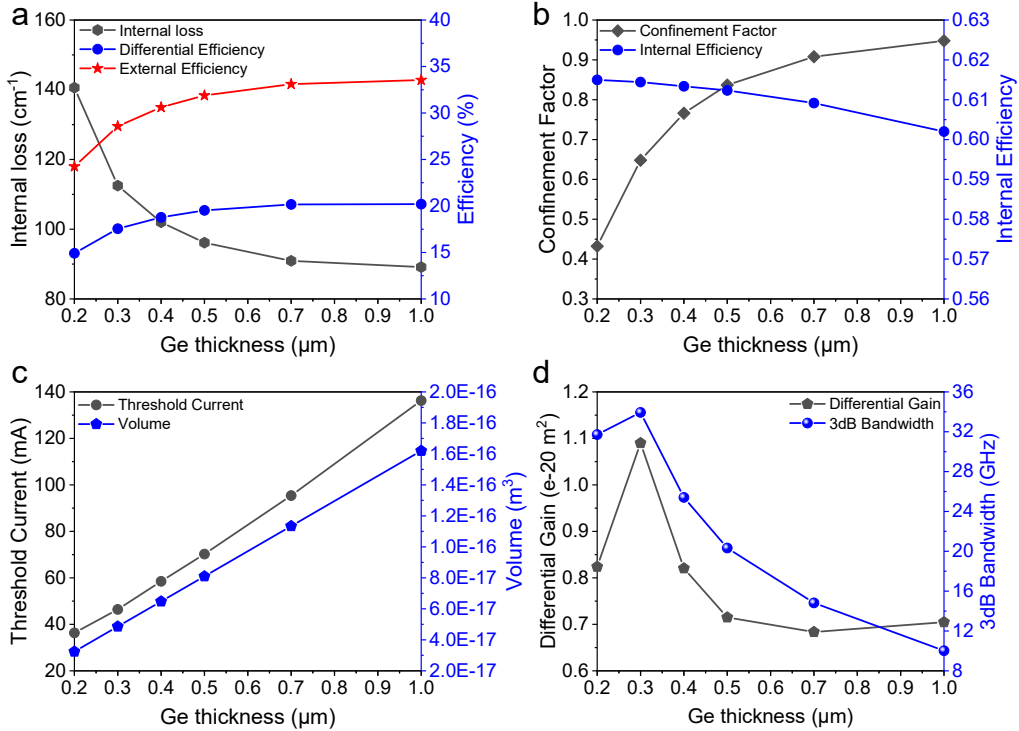

Figure 7. Ge thickness dependence in the range of 0.2 to 1.0 μm ($d_{poly-Si}$ = 0.7 μm, $W_{Ge}$ = 0.6 μm). a) The internal loss, $\langle \alpha_i \rangle$, differential efficiency, $\eta_d$ and the extraction efficiency, $\eta_{ext}$. b) Confinement factor, $\Gamma$, and internal efficiency, $\eta_i$. c) Threshold current, $I_{th}$, and volume of the Ge active region. d) The differential gain, $dg/dN$, and 3dB bandwidth at the biased current of 1002 A/m or 270.5 mA. This biased current value was chosen to be larger than the threshold current of all geometry changing range, and not higher than 10 times $I_{th}$ to guarantee the lasers working properly.

### 4.4 Defect-limited minority carrier lifetime dependence

In the previous optimization, the defect limited minority lifetime ($\tau_{n,p}$) was set as 1 ns for conservative estimation while the minority carrier lifetime strongly depends on the concentration of recombination centers and can be utilized to determine the Shockley-Read-Hall (SRH) recombination rate, $R_{SRH}$ [45].

$$R_{SRH} = \sigma_{n,p} v_{n,p} N_t \delta n, \qquad (12)$$

where $\sigma_{n,p}$ is the electron and hole capture cross sections of deep traps, $v_{n,p}$ is the thermal velocity of electrons and holes, $N_t$ is the trap (or defect) density, $\delta n$ is the excessive electron concentration. The capture coefficient $c_{p,n}$ for electrons and holes is related to the lifetime of the carrier by following equations [45]:

$$\frac{1}{\tau_{n,p}} = c_{n,p} N_t, \tag{13}$$

$$c_{n,p} = \sigma_{n,p} v_{n,p}, \tag{14}$$

where $N_t$ is the trap density, $\sigma_{n,p}$ is the electron and hole capture cross sections of deep traps, $v_{n,p}$ is the thermal velocity of electrons and holes. To obtain Ge-on-Si layers with higher quality and longer minority lifetime can be realized by growing Ge on a GOI substrates or directly wafer bonding [49] and chemical mechanical polishing (CMP) [50]. Researchers have reported minority lifetime of 3.12 ns and 5.3 ns of Ge layers using these strategies [49, 50]. Thin film delamination from bulk Ge wafers, like smart cut technology in Si, may also be able to provide higher quality Ge thin films.

The effect of minority lifetime and possible defect density range have been investigated based on the optimized geometrical structure and only changing $\tau_{n,p}$, listed in Table II. From these results, we can tell that by enhancing the Ge materials quality, the laser performance can be improved. With longer minority carrier lifetime, the carriers can stay longer and recombine slower in the cavity, which means the injection carrier is less needed for the lasers and therefore decrease the threshold current. As for the threshold current, it can be reduced by 6.7 times by increasing the minority lifetime from 1 ns to 10 ns, and 15 times when increasing to 100 ns. The differential efficiency and differential gain show no variation with the change of minority lifetime, because this only influences the SRH recombination rate, $R_{SRH}$, which affects the threshold current but does not change the internal loss or optical confinement factor. The modulation bandwidth depends on the square root of the relative value of biased current to the threshold current value. Under this condition, the 3dB bandwidth increases slightly with the improvement of minority lifetime.

Table II The laser performance with different minority lifetime

| Minority lifetime (typical dislocation density) | Threshold Current (mA) | Differential efficiency (%) | Differential Gain ($m^2$) | 3dB bandwidth (GHz) |
|---|---|---|---|---|
| 1 ns ($1 \times 10^7$ cm$^{-2}$ [27]) | 46.42 | 17.6 | $1.09 \times 10^{-20}$ | 33.94 |
| 10 ns | 6.85 | 17.6 | $1.09 \times 10^{-20}$ | 36.89 |
| 100 ns ($1 \times 10^5$ cm$^{-2}$ [27]) | 2.92 | 17.6 | $1.09 \times 10^{-20}$ | 37.01 |

### 4.5 Eye diagrams of the optimized laser structure

An eye diagram is a useful tool in visualizing intersymbol interference between data bits and diagnosing communication link problems. Therefore, based on the above optimized laser structure, we also predicted the digital modulation property of our optimized laser device ($d_{poly-Si}$ = 0.7 μm, $W_{Ge}$ = 0.6 μm, $d_{Ge}$ = 0.3 μm), and the minority lifetime $\tau_{n,p}$ = 1 ns, shown in Figure 8. The transmission bit rates varied from 10 to 40 Gb/s, which is in a back-to-back (BTB) configuration with an extinction ratio of 3.44 dB at a biased current of 270.5 mA. With a 10 and 20 Gb/s non-return-to-zero (NRZ) signal, a clear eye opening of the optical signal is obtained as seen in Figure 8(a) and (b). Furthermore, when the bit rate is raised up to 32 and 40 Gb/s, eye-opening window narrows, as shown in Figure 8(c) and (d).

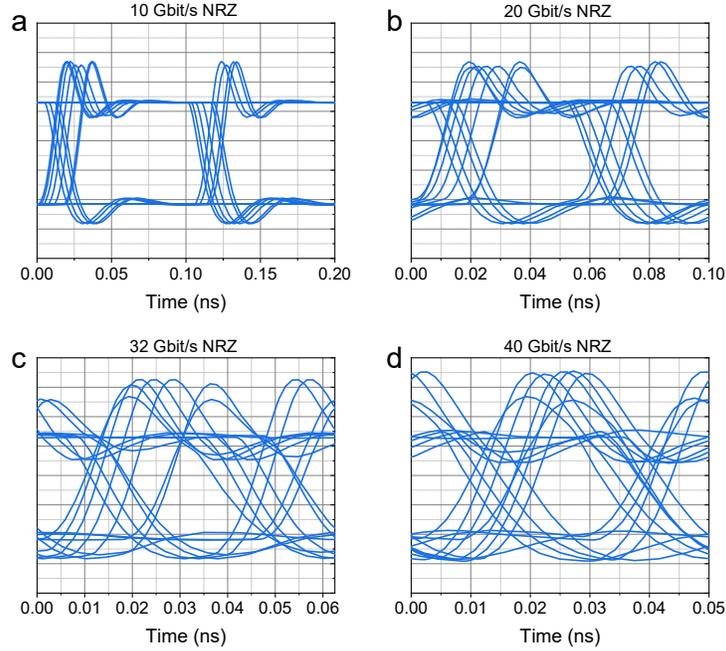

Figure 8. Simulated eye diagrams for the optimized structure ($d_{poly-Si}$ = 0.7 μm, $W_{Ge}$ = 0.6 μm, $d_{Ge}$ = 0.3 μm, $\tau_{n,p}$ = 1 ns) at different bit rates. a) 10 Gb/s NRZ. b) 20 Gb/s NRZ. c) 32 Gb/s NRZ. d) 40 Gb/s NRZ.

## 5 Conclusion

In this work, LASTIP™ was used to model and simulate the small signal modulation responses of the Fabry-Perot Ge-on-Si laser diodes. The geometrical parameters, such as poly-Si cladding thickness, Ge cavity width and thickness were studied and optimized for better 3dB bandwidth. A threshold current of 46.42 mA and a 3dB bandwidth of 33.94 GHz at a biased current of 270.5 mA were predicted with an optimized laser structure, where $d_{poly-Si}$ = 0.7 μm, $W_{Ge}$ = 0.6 μm and $d_{Ge}$ = 0.3 μm with 1 ns minority carrier lifetime. The eye diagrams simulated show a stable eye-opening window at 20 Gb/s NRZ. The improvement to 10 ns minority carrier lifetime would reduce the threshold current to 6.85 mA, and increases the 3dB bandwidth to 36.89 GHz.

Better Ge material quality, strain and doping refinement and improving the differential gain are all important to enhance the Ge-on-Si laser performance. Our work paved the way for further improvement of Ge lasers and shed light on the silicon integrated optoelectronic devices.

## Acknowledgement
The authors would like to thank Crosslight Software, Inc., for providing the license of LASTIP™. Dr. Rodolfo Camacho-Aguilera at Luminous Computing is acknowledged for helpful discussions and proofreading.

**Reference**

[1] Z. Zhou, B. Yin, and J. Michel, "On-chip light sources for silicon photonics," *Light: Science & Applications,* vol. 4, no. 11, pp. e358-e358, 2015, doi: 10.1038/lsa.2015.131.

[2] C. R. Doerr, "Silicon photonic integration in telecommunications," *Frontiers in Physics,* vol. 3, 2015, doi: 10.3389/fphy.2015.00037.

[3] J. Wang and Y. Long, "On-chip silicon photonic signaling and processing: a review," *Science Bulletin,* vol. 63, no. 19, pp. 1267-1310, 2018, doi: 10.1016/j.scib.2018.05.038.

[4] P. Dong, Y.-K. Chen, G.-H. Duan, and D. T. Neilson, "Silicon photonic devices and integrated circuits," *Nanophotonics,* vol. 3, no. 4-5, pp. 215-228, 2014, doi: 10.1515/nanoph-2013-0023.

[5] Y. Liu, S. Wang, J. Wang, X. Li, M. Yu, and Y. Cai, "Silicon photonic transceivers in the field of optical communication," *Nano Communication Networks,* vol. 31, 2022, doi: 10.1016/j.nancom.2021.100379.

[6] A. W. Fang, H. Park, O. Cohen, R. Jones, M. J. Paniccia, and J. E. Bowers, "Electrically pumped hybrid AlGaInAs-silicon evanescent laser," *Opt Express,* vol. 14, no. 20, pp. 9203-9210, 2006.

[7] J. Liu, "Monolithically Integrated Ge-on-Si Active Photonics," *Photonics,* vol. 1, no. 3, pp. 162-197, 2014, doi: 10.3390/photonics1030162.

[8] S. M. Chen *et al.*, "1.3 μm InAs/GaAs quantum‐dot laser monolithically grown on Si substrates operating over 100°C," *Electronics Letters,* vol. 50, no. 20, pp. 1467-1468, 2014, doi: 10.1049/el.2014.2414.

[9] B. Szelag *et al.*, "Hybrid III–V/Silicon Technology for Laser Integration on a 200-mm Fully CMOS-Compatible Silicon Photonics Platform," *IEEE Journal of Selected Topics in Quantum Electronics,* vol. 25, no. 5, pp. 1-10, 2019, doi: 10.1109/jstqe.2019.2904445.

[10] S. Yerci, R. Li, and L. Dal Negro, "Electroluminescence from Er-doped Si-rich silicon nitride light emitting diodes," *Applied Physics Letters,* vol. 97, no. 8, 2010, doi: 10.1063/1.3483771.

[11] H. Kataria *et al.*, "Simple Epitaxial Lateral Overgrowth Process as a Strategy for Photonic Integration on Silicon," *IEEE Journal of Selected Topics in Quantum Electronics,* vol. 20, no. 4, pp. 380-386, 2014, doi: 10.1109/jstqe.2013.2294453.

[12] M. E. Groenert *et al.*, "Monolithic integration of room-temperature cw GaAs/AlGaAs lasers on Si substrates via relaxed graded GeSi buffer layers," *Journal of Applied Physics,* vol. 93, no. 1, pp. 362-367, 2003, doi: 10.1063/1.1525865.

[13] X. Sun *et al.*, "Electrically pumped hybrid evanescent Si/InGaAsP lasers," *Optics Letters,* vol. 34, no. 9, pp. 1345-1347, 2009.

[14] S. Keyvaninia, M. Muneeb, S. Stanković, P. J. V. Veldhoven, D. V. Thourhout, and G. Roelkens, "Ultra-thin DVS-BCB adhesive bonding of III-V wafers, dies and multiple dies to a patterned silicon-on-insulator substrate," *Opt Mater Express,* vol. 3, no. 1, pp. 35-46, 2013.

[15] A. Y. Liu *et al.*, "High performance continuous wave 1.3 μm quantum dot lasers on silicon," *Applied Physics Letters,* vol. 104, no. 4, 2014, doi: 10.1063/1.4863223.

[16] S. Chen *et al.*, "Electrically pumped continuous-wave III–V quantum dot lasers on silicon," *Nature Photonics,* vol. 10, no. 5, pp. 307-311, 2016, doi: 10.1038/nphoton.2016.21.

[17] A. Lee, Q. Jiang, M. Tang, A. Seeds, and H. Liu, "Continuous-wave InAs/GaAs quantum-dot laser diodes monolithically grown on Si substrate with low threshold current densities," *Optics Letters,* vol. 34, no. 20, pp. 22181-22187, 2012.

[18] G. Z. Mashanovich *et al.*, "Germanium Mid-Infrared Photonic Devices," *Journal of Lightwave Technology,* vol. 35, no. 4, pp. 624-630, 2017, doi: 10.1109/jlt.2016.2632301.

[19]  V. Reboud *et al.*, "Germanium based photonic components toward a full silicon/germanium photonic platform," *Progress in Crystal Growth and Characterization of Materials,* vol. 63, no. 2, pp. 1-24, 2017, doi: 10.1016/j.pcrysgrow.2017.04.004.

[20]  F. T. Armand Pilon *et al.*, "Lasing in strained germanium microbridges," *Nat Commun,* vol. 10, no. 1, p. 2724, Jun 20 2019, doi: 10.1038/s41467-019-10655-6.

[21]  Y. Lin *et al.*, "High-efficiency normal-incidence vertical p-i-n photodetectors on a germanium-on-insulator platform," *Photonics Research,* vol. 5, no. 6, p. 702, 2017, doi: 10.1364/prj.5.000702.

[22]  O. Madelung, O. Madelung, Ed. *Semiconductors Group IV Elements and III-V Compounds*. 1991.

[23]  J. Liu *et al.*, "Tensile-strained, n-type Ge as a gain medium for monolithic laser integration on Si," (in English), *Opt Express,* vol. 15, no. 18, pp. 11272-11277, Sep 3 2007, doi: Doi 10.1364/Oe.15.011272.

[24]  J. Liu, X. Sun, R. Camacho-Aguilera, L. C. Kimerling, and J. Michel, "Ge-on-Si laser operating at room temperature," (in English), *Opt Lett,* vol. 35, no. 5, pp. 679-81, Mar 1 2010, doi: 10.1364/OL.35.000679.

[25]  R. E. Camacho-Aguilera *et al.*, "An electronically pumped Ge laser," *Opt Express,* 2012.

[26]  R. Koerner *et al.*, "Electrically pumped lasing from Ge Fabry-Perot resonators on Si," *Opt Express,* vol. 23, no. 11, pp. 14815-22, Jun 1 2015, doi: 10.1364/OE.23.014815.

[27]  R. E. Camacho-Aguilera, "Ge-on-Si LASER for Silicon Photonics," Doctor of Philosophy in Materials Science and Engineering, Materials Science & Engineering Georgia Institute of Technology, Massachusetts Institute of Technology, 2013.

[28]  J. Jiang and J. Sun, "Theoretical analysis of optical gain in uniaxial tensile strained and $n^+$-doped Ge/GeSi quantum well," *Opt Express,* vol. 24, no. 13, pp. 14525-37, Jun 27 2016, doi: 10.1364/OE.24.014525.

[29]  J. Ke, L. Chrostowski, and G. Xia, "Stress Engineering With Silicon Nitride Stressors for Ge-on-Si Lasers," *IEEE Photonics Journal,* vol. 9, no. 2, pp. 1-15, 2017, doi: 10.1109/jphot.2017.2675401.

[30]  J. Petykiewicz *et al.*, "Direct Bandgap Light Emission from Strained Germanium Nanowires Coupled with High-Q Nanophotonic Cavities," *Nano Lett,* vol. 16, no. 4, pp. 2168-73, Apr 13 2016, doi: 10.1021/acs.nanolett.5b03976.

[31]  S. Bao *et al.*, "Low-threshold optically pumped lasing in highly strained germanium nanowires," *Nat Commun,* vol. 8, no. 1, p. 1845, Nov 29 2017, doi: 10.1038/s41467-017-02026-w.

[32]  P. A. Morton *et al.*, "25 GHz bandwidth 1.55 µm GaInAsP p-doped strained multiquantum-well lasers," *Electronics Letters,* vol. 28, no. 23, pp. 2156 – 2157, 1992, doi: 10.1049/el:19921384.

[33]  W. Kobayashi *et al.*, "50-Gb/s Direct Modulation of a 1.3-µm InGaAlAs-Based DFB Laser With a Ridge Waveguide Structure," *IEEE Journal of Selected Topics in Quantum Electronics,* vol. 19, no. 4, pp. 1500908-1500908, 2013, doi: 10.1109/jstqe.2013.2238509.

[34]  S. Golovynskyi *et al.*, "Interband Photoconductivity of Metamorphic InAs/InGaAs Quantum Dots in the 1.3-1.55-mum Window," *Nanoscale Res Lett,* vol. 13, no. 1, p. 103, Apr 16 2018, doi: 10.1186/s11671-018-2524-3.

[35]  S. Yamaoka *et al.*, "Directly modulated membrane lasers with 108 GHz bandwidth on a high-thermal-conductivity silicon carbide substrate," *Nature Photonics,* vol. 15, no. 1, pp. 28-35, 2020, doi: 10.1038/s41566-020-00700-y.

[36] X. Li, Z. Li, S. Li, L. Chrostowski, and G. Xia, "Design considerations of biaxially tensile-strained germanium-on-silicon lasers," *Semiconductor Science and Technology,* vol. 31, no. 6, p. 065015, 2016, doi: 10.1088/0268-1242/31/6/065015.

[37] D. S. Sukhdeo, S. Gupta, K. C. Saraswat, B. Dutt, and D. Nam, "Impact of minority carrier lifetime on the performance of strained germanium light sources," (in English), *Optics Communications,* vol. 364, pp. 233-237, Apr 1 2016, doi: 10.1016/j.optcom.2015.11.060.

[38] R. Newman and W. W. Tyler, "Effect of Impurities on Free-Hole Infrared Absorption in p-Type Germanium," *Physical Review,* vol. 105, no. 3, pp. 885-886, 1957, doi: 10.1103/PhysRev.105.885.

[39] D. K. Schroder, R. N. Thomas, and J. C. Swartz, "Free Carrier Absorption in Silicon," *IEEE Journal of Solid-State Circuits,* vol. 13, no. 1, pp. 180-187, 1978.

[40] O. Ogah, "Free-carrier effects in polycrystalline silicon-on-insulator photonic devices," Master of Science, Microelectronic Engineering, ROCHESTER INSTITUTE OF TECHNOLOGY, 2010.

[41] L. A. Coldren, S. W. Corzine, and M. L. Masanovic, K. Chang, Ed. *Diode Lasers and Photonic Integrated Circuits*, Second Edition ed. John Wiley & Sons, Inc., Hoboken, New Jersey, 2012.

[42] C. C. Shen *et al.*, "Design, Modeling, and Fabrication of High-Speed VCSEL with Data Rate up to 50 Gb/s," *Nanoscale Res Lett,* vol. 14, no. 1, p. 276, Aug 14 2019, doi: 10.1186/s11671-019-3107-7.

[43] E. Kapon, E. Kapon, Ed. *Semiconductor Lasers I. Fundamentals*. Academic Press, 1999.

[44] N. Hatori, M. Sugawara, K. Mukai, Y. Nakata, and H. Ishikawa, "Room-temperature gain and differential gain characteristics of self-assembled InGaAs/GaAs quantum dots for 1.1–1.3 μm semiconductor lasers," *Applied Physics Letters,* vol. 77, no. 6, pp. 773-775, 2000, doi: 10.1063/1.1306662.

[45] S. L. Chuang, *Physics of Photonic Devices*, Second ed. New Jersey: John Wiley & Sons, 2009.

[46] *Crossligh LASTIP. Available: https://crosslight.com/products/lastip/.* (2020).

[47] J. HUANG and L. W. CASPERSON, "Gain and saturation in semiconductor lasers," *Optical and Quantum Electronics,* vol. 25, pp. 369-390, 1993.

[48] Y. Cai, "Materials Science and Design for Germanium Monolithic Light Source on Silicon," Doctor of Philosophy in Materials Science and Engineering, Department of Materials Science and Engineering, Massachusetts Institute of Technology, 2014.

[49] R. Geiger *et al.*, "Excess carrier lifetimes in Ge layers on Si," *Applied Physics Letters,* vol. 104, no. 6, 2014, doi: 10.1063/1.4865237.

[50] D. Nam, J. H. Kang, M. L. Brongersma, and K. C. Saraswat, "Observation of improved minority carrier lifetimes in high-quality Ge-on-insulator using time-resolved photoluminescence," *Opt Lett,* vol. 39, no. 21, pp. 6205-8, Nov 1 2014, doi: 10.1364/OL.39.006205.